\pdfoutput=1

\documentclass[11pt,nofootinbib]{article}
\usepackage{amsmath,amssymb,mathtools}
\usepackage{graphicx}
\usepackage[colorlinks=true,linkcolor=blue,citecolor=blue,urlcolor=blue,linktocpage=true]{hyperref}
\usepackage{subfigure}
\usepackage{comment}
\usepackage{tensor}
\usepackage{cases}
\usepackage[utf8]{inputenc}
\usepackage{cite}
\usepackage[font=small,labelfont=bf]{caption}

\numberwithin{equation}{section}

\begin{document}

\providecommand{\abs}[1]{\lvert#1\rvert}
\providecommand{\bd}[1]{\boldsymbol{#1}}

\begin{titlepage}

\setcounter{page}{1} \baselineskip=15.5pt \thispagestyle{empty}

\begin{flushright}
\end{flushright}
\vfil

\bigskip
\begin{center}
 {\LARGE \textbf{Axionlike Origin of the}}\\
 \medskip
 {\LARGE \textbf{Primordial Density Perturbation}}
\vskip 15pt
\end{center}

\vspace{0.5cm}
\begin{center}
{\large
Takeshi Kobayashi
}\end{center}

\vspace{0.3cm}

\begin{center}
\textit{Kobayashi-Maskawa Institute for the Origin of Particles and the Universe,\\ Nagoya University, Nagoya 464-8602, Japan}\\

\vskip 14pt
E-mail:
 \texttt{\href{mailto:takeshi@kmi.nagoya-u.ac.jp}{takeshi@kmi.nagoya-u.ac.jp}}
\end{center} 



\vspace{1cm}

\noindent
We show that an axionlike field coupled to a new confining gauge group can generate the primordial density perturbation in the post-inflation universe. The axion decay constant and strong coupling scale are uniquely determined by observations of the density perturbation, which further suggest a temporal deconfinement of the gauge group after inflation. The resulting temperature-dependent axion potential, together with its periodic nature, gives rise to the red-tilted density perturbation, with a positive local-type non-Gaussianity of order unity.
\vfil

\end{titlepage}

\newpage

\section{Introduction}

The large-scale structure of our universe was seeded by tiny
density fluctuations in the early universe.
This picture of structure formation has been established by a variety of
cosmological observations, which also revealed that the
primordial density perturbation was adiabatic, nearly 
scale-invariant with a slight increase of the amplitude towards larger
length scales (red-tilted), and nearly Gaussian.
However, the physical origin of the density perturbation still remains
a mystery.
One possibility is that it was produced during cosmic inflation
from fluctuations of the inflaton field.
Alternatively, it could have been produced after inflation by another
physical degree of freedom.

In theories beyond the standard model of particle physics, the
appearance of pseudo-Nambu-Goldstone bosons (PNGBs) of spontaneously 
broken global U(1) symmetries is ubiquitous.
The most famous example is the QCD axion introduced for solving
the strong $CP$ problem~\cite{Peccei:1977hh}.
Another well-studied example occurs in string theory, which suggests the
existence of many such particles in the four-dimensional effective
theory upon string compactifications~\cite{Svrcek:2006yi}. 
Axions and axionlike particles can also play important roles in
cosmology, and in particular the possibility that they constitute the
dark matter of our universe has been extensively studied.

In this letter we argue that an axionlike field can serve as a
curvaton~\cite{Mollerach:1989hu} and generate the primordial density
perturbation in the post-inflation universe. 
Considering the axion to be coupled to a new gauge force that becomes
strong at low energies, 
we show that the observed characteristics of the density perturbation
require the strong coupling scale to be higher than the de Sitter
temperature of the inflationary universe, but lower than the maximum
radiation temperature during reheating, i.e.,
$ T_{\mathrm{inf}} \ll \Lambda \ll T_{\mathrm{max}}$.
This leads to a temporal deconfinement of the gauge group after
inflation, with which the axion generates the density perturbation 
by virtue of its temperature-dependent and periodic potential.

Our scenario is quite distinct from previous studies on 
PNGB curvatons~\cite{Dimopoulos:2003az,Kawasaki:2011pd}.
For a curvaton to produce the observed red 
spectral index $n_s -1 \sim -10^{-2}$,
one needs to require either 
(i) a time derivative of the Hubble rate during inflation of
$-\dot{H}/H^2 \sim 10^{-2}$, and/or
(ii) a tachyonic curvaton mass with amplitude
$\abs{m} \sim 10^{-1} H_{\mathrm{inf}}$ in terms of the inflationary
Hubble rate. Option~(i) implies that the inflaton field travels a
super-Planckian distance, and thus imposes strong
constraints on inflationary model building. 
In option~(ii), the rather large tachyonic mass forces the curvaton to
start oscillating soon after inflation ends, which in turn makes it
challenging for the curvaton to dominate the post-inflation
universe and produce the density perturbation.
One way to avoid the early onset of the oscillation is to place
the curvaton at a fine-tuned initial position close to a potential
maximum, however this also enhances the non-Gaussianity and thus
is ruled out by current observations. 
We will show that an axionlike field coupled to a strong sector
can generate a red-tilted and nearly Gaussian perturbation via (ii), by
taking advantage of a temperature-dependent potential which naturally
delays the oscillation, without the need of fine-tuning the initial
condition.

\section{The axion setup}

We consider a PNGB of some global U(1) symmetry that is
spontaneously broken at an energy scale~$f$; we refer to this field as
the axion and $f$ the axion decay constant. 
We also assume the U(1) to be explicitly broken due to the axion
being coupled to a new gauge force (not QCD) that becomes strong at low
energies, yielding a periodic axion potential. 
Considering the periodicity to be governed by $f$, 
we write the axion Lagrangian as
\begin{equation}
-\frac{1}{2} f^2 \partial_\mu \theta \partial^\mu \theta
- m(T)^2 f^2 \left( 1 - \cos \theta \right)  
+ \frac{ \alpha }{8 \pi } \theta F_{\mu \nu} \tilde{F}^{\mu \nu }.
\label{eq:L}
\end{equation}
Here the axion is written as a dimensionless angle~$\theta$
with periodicity $\theta \cong \theta + 2 \pi$,
and we have also added a pseudoscalar coupling to photons (or hidden
photons).
The axion potential arises below some strong coupling scale~$\Lambda$
($ <  f$), and we consider the axion mass to depend on the cosmic
temperature as  
\begin{equation}
 m(T) \simeq 
 \begin{dcases}
    m_0 \left(\frac{\Lambda}{T}\right)^p
       & \mathrm{for}\, \, \, T > \Lambda , \\
    m_0 
       & \mathrm{for}\, \, \, T < \Lambda,
 \end{dcases}
\label{eq:mofT}
\end{equation}
with a positive power $p$ of order unity, although our main results are
insensitive to the detailed form of the temperature dependence
above~$\Lambda$. 
The zero-temperature axion mass is given by
\begin{equation}
 m_0 = \frac{\Lambda^2}{f}.
\end{equation}
Without loss of generality we take the axion to lie within the range
$-\pi \leq \theta \leq \pi$.

\section{Cosmological evolution}

Upon analyzing the axion dynamics, we will
make a couple of assumptions along the way.
We will later show that most of the assumptions are actually required by
the observed characteristics of the density perturbation.

We first assume that the de Sitter temperature during inflation, 
$T_{\mathrm{inf}} = H_{\mathrm{inf}} / 2 \pi $,
is lower than the strong coupling scale, i.e.,
\begin{equation}
 T_{\mathrm{inf}} < \Lambda < f.
\end{equation}
Hence during inflation the U(1) is already broken,
and moreover the axion potential takes its zero-temperature form.
Further supposing $m_0^2 \ll H_{\mathrm{inf}}^2$, and that
$f$ is smaller than the reduced Planck scale
$M_{\mathrm{Pl}} = (8 \pi G)^{-1/2}$,
then $\Lambda^4 \ll M_{\mathrm{Pl}}^2 H_{\mathrm{inf}}^2$,
i.e., the energy density of the axion is tiny compared to the total
density of the universe 
and thus the axion serves as a spectator.
Supposing that the axion evolution is dominated by classical rolling
and not quantum fluctuations, the axion slowly rolls as
$3 H_{\mathrm{inf}} \dot{\theta} \simeq -m_0^2 \sin \theta$. 
Integrating this by ignoring any time dependence of the Hubble rate
yields 
\begin{equation}
 \tan \left( \frac{\theta_{\mathrm{end}}}{2}  \right) \simeq
 \tan \left( \frac{\theta_1}{2}\right) 
\cdot \exp \left( - \frac{m_0^2}{3 H_{\mathrm{inf}}^2}\mathcal{N}_1
  \right).
\label{eq:tan_theta}
\end{equation}
Here $\theta_{\mathrm{end}}$ is the field value when inflation ends,
and $\mathcal{N}_1$ is the number of $e$-folds since the time when the
axion takes a value~$\theta_1$ until the end of inflation. 

After inflation, the inflaton is considered to decay and
trigger a radiation-dominated epoch.
We assume that the maximum value of the radiation temperature lies
within the range
\begin{equation}
 \Lambda < T_{\mathrm{max}} < f,
\label{eq:i}
\end{equation}
and thus the U(1) symmetry stays broken, however the axion potential
diminishes. 
The energy density and temperature of radiation
are related via
$\rho_{\mathrm{r}} = (\pi^2/30) g_* T^4$,
and for simplicity 
we take the number of relativistic degrees of freedom as a constant.
As a minimal model, we consider the inflaton to instantaneously decay
at the end of inflation; then the maximum radiation temperature is
\begin{equation}
 T_{\mathrm{max}} = 
\left( \frac{90}{\pi^2 g_*}    \right)^{1/4}
(M_{\mathrm{Pl}} H_{\mathrm{inf}})^{1/2}.
\label{eq:T_max}
\end{equation}
The axion slow-rolls until its potential diminishes, 
after which it freely streams for a few Hubble times and then
comes to a halt due to the Hubble friction.
The field excursion during the free-streaming can be estimated 
using the slow-roll approximation just before the end of inflation as
$\abs{\Delta \theta } \sim
\abs{\dot{\theta}_{\mathrm{end}}/ H_{\mathrm{inf}}  }
< (m_0^2 / H_{\mathrm{inf}}^2) \abs{\theta_{\mathrm{end}}} \ll
\abs{\theta_{\mathrm{end}}} $. 
This shows that the axion is effectively frozen at 
$\theta_{\mathrm{end}}$ 
while its potential is negligible.

The axion potential arises again as the universe cools down.
When the mass becomes comparable to the Hubble rate, the axion
begins to oscillate about its potential minimum, 
and thereafter the axion number is conserved.
The temperature when the oscillation begins can be computed, supposing
it is above $\Lambda$ and that the universe is dominated by radiation, as 
\begin{equation}
 T_{\mathrm{osc}} \simeq \Lambda 
\left( \frac{\sqrt{90}}{\pi g_*^{1/2}  c}
\frac{M_{\mathrm{Pl}}}{f} \right)^{\frac{1}{p+2}}.
\end{equation}
The subscript ``osc'' denotes quantities at the onset of the
oscillation, and 
$c \equiv m(T_{\mathrm{osc}}) / H_{\mathrm{osc}}$
is the mass-to-Hubble ratio at this time.
Here, note that $ m(T_{\mathrm{osc}}) < m_0 $.
The number density of the axion is written as
$n_\theta = \rho_\theta / m(T)$
in terms of the energy density 
$\rho_\theta = f^2 \dot{\theta}^2/2 + m(T)^2 f^2 (1-\cos \theta)$.
Hence its value at the onset of the oscillation is
\begin{equation}
 n_{\theta \mathrm{osc}} \simeq m(T_{\mathrm{osc}}) f^2
\left(1 - \cos \theta_{\mathrm{end}} \right),
\label{eq:n_theta}
\end{equation}
and thereafter redshifts as $n_\theta \propto a^{-3}$.

Since the radiation density redshifts as $\rho_{\mathrm{r}} \propto
a^{-4}$, the axion would eventually dominate the universe if it is
sufficiently long-lived.
The Hubble rate when axion domination takes over,
i.e., when $\rho_\theta = \rho_{\mathrm{r}} $,
is obtained as 
\begin{equation}
 H_{\mathrm{dom}} \simeq \frac{60 \sqrt{5}}{\pi^3 g_*^{3/2}}
\frac{\Lambda^2}{M_{\mathrm{Pl}}}
\left( \frac{ \Lambda }{T_{\mathrm{osc}}} \right)^{2p+6}
\left(1 - \cos \theta_{\mathrm{end}} \right)^2,
\end{equation}
where we have considered the temperature 
at this time to satisfy 
$T_{\mathrm{dom}} < \Lambda$ so that 
$\rho_\theta = m_0 n_\theta $.
Whether the axion actually dominates the universe
depends on its lifetime.
At temperatures below $\Lambda$, the decay width of the axion induced by 
the axion-photon coupling in (\ref{eq:L}) is
\begin{equation}
 \Gamma = \frac{\alpha^2}{256 \pi^3} \frac{\Lambda^6}{f^5}.
\end{equation}
If $\Gamma < (>) \, H_{\mathrm{dom}}$, then 
the axion would decay after (before) dominating the universe.
Supposing that the axion decays suddenly when $H = \Gamma$
(we denote quantities at this time by the subscript ``dec''),
one finds a relation:
\begin{equation}
 3 M_{\mathrm{Pl}}^2 \Gamma^2 = 
 3 M_{\mathrm{Pl}}^2 H_{\mathrm{osc}}^2
 \left(\frac{a_{\mathrm{osc}}}{a_{\mathrm{dec}}}\right)^4 
+ m_0 n_{\theta \mathrm{osc}}
\left(\frac{a_{\mathrm{osc}}}{a_{\mathrm{dec}}}\right)^3 .
\label{eq:sum}
\end{equation}
The terms in the right hand side correspond to the radiation
and axion densities right before the decay,
and we write their ratio as
$R \equiv (\rho_{\theta} / \rho_{\mathrm{r}} )_{\mathrm{dec}}$
for later convenience.

\section{Curvature perturbation}

The axion acquires super-horizon field fluctuations during inflation,
which later get converted into cosmological density (or curvature)
perturbations as the axion comes to dominate the universe. 
The generation of the perturbation completes when the axion decays. 
Using the $\delta \mathcal{N}$~formalism~\cite{Starobinsky:1986fxa},
we compute the axion-induced curvature perturbation as the
fluctuation in the number of $e$-folds~$\mathcal{N}$ between an initial
flat slice during inflation when a comoving wave number~$k$ of
interest exists the horizon ($k = a H$),
and a final uniform-density slice where $H = \Gamma$. 
Noting that the axion field fluctuations on the initial slice is
Gaussian with a power spectrum $P_{\delta \theta } (k) \simeq
(H_k / 2 \pi f)^2$
(the subscript ``$k$'' refers to quantities when $k = a H$), 
the power spectrum of the curvature perturbation is written as 
$P_\zeta (k) \simeq (\partial \mathcal{N} / \partial \theta_k)
(H_k / 2 \pi f)^2$.
Likewise, the non-Gaussianity parameter characterizing the
amplitude of the local bispectrum is 
$f_{\mathrm{NL}} \simeq (5/6) 
(\partial^2 \mathcal{N} / \partial \theta_k^2) 
(\partial \mathcal{N} / \partial \theta_k)^{-2} $.

The derivatives of~$\mathcal{N} = \ln (a_{\mathrm{dec}} / a_k)$ can be
evaluated by differentiating (\ref{eq:sum}) multiple times with respect
to $\theta_k$. 
In the vicinity of $\theta = 0$ where the axion potential is
well-approximated by a quadratic, the mass-to-Hubble ratio~$c$ at the
onset of the oscillation
is independent of the axion field value~\cite{quadratic_c};
however this is not the case in the region close to the hilltop
$\abs{\theta} = \pi$~\cite{Kawasaki:2011pd}.
Let us for the moment suppose that 
$\theta_{\mathrm{end}}$ is not too far from $0$
and take $\partial c / \partial \theta_{k}  = 0$.
Then, also because the axion density is negligibly tiny 
compared to the total density of the universe during
$a_k \leq a \leq a_{\mathrm{osc}}$,
the ratio $(a_{\mathrm{osc}} / a_k)$ is independent of
$\theta_k$. 
Derivatives of $\theta_{\mathrm{end}}$ can be evaluated
using~(\ref{eq:tan_theta}).  
Hence, after some manipulation we find
\begin{equation}
\begin{split}
& P_\zeta (k) \simeq
\left( \frac{R}{3 R + 4} 
\frac{1 + \cos \theta_{\mathrm{end}}}{\sin \theta_k}\frac{H_k}{2 \pi f}
\right)^2,
\\
& f_{\mathrm{NL}} \simeq
\frac{5}{6} \left\{
\frac{4}{R} + \frac{4}{3 R + 4}
-  \left( 3 + \frac{4}{R} \right)
\frac{1 - \cos \theta_{\mathrm{end}} + \cos \theta_k}{1 + \cos
\theta_{\mathrm{end}}} 
\right\}.
\end{split}
\label{eq:Pzeta-fNL}
\end{equation}
In the $\theta_k \to 0$ limit, 
these expressions reduce to those of the vanilla curvaton with a quadratic
potential. 
The spectral index of the power spectrum 
$n_s - 1 = d \ln P_\zeta / d \ln k$ 
is computed using $d \ln k \simeq H_k d t$ and the slow-roll
approximation during inflation as
\begin{equation}
 n_s - 1 \simeq 
\frac{2}{3} \frac{m_0^2 }{H_k^2} \cos \theta_k
+ 2  \frac{\dot{H}_k}{H_k^2}.
\label{eq:n_s-1}
\end{equation}

\begin{figure*}[t]
 \begin{minipage}{.32\linewidth}
  \begin{center}
 \includegraphics[width=\linewidth]{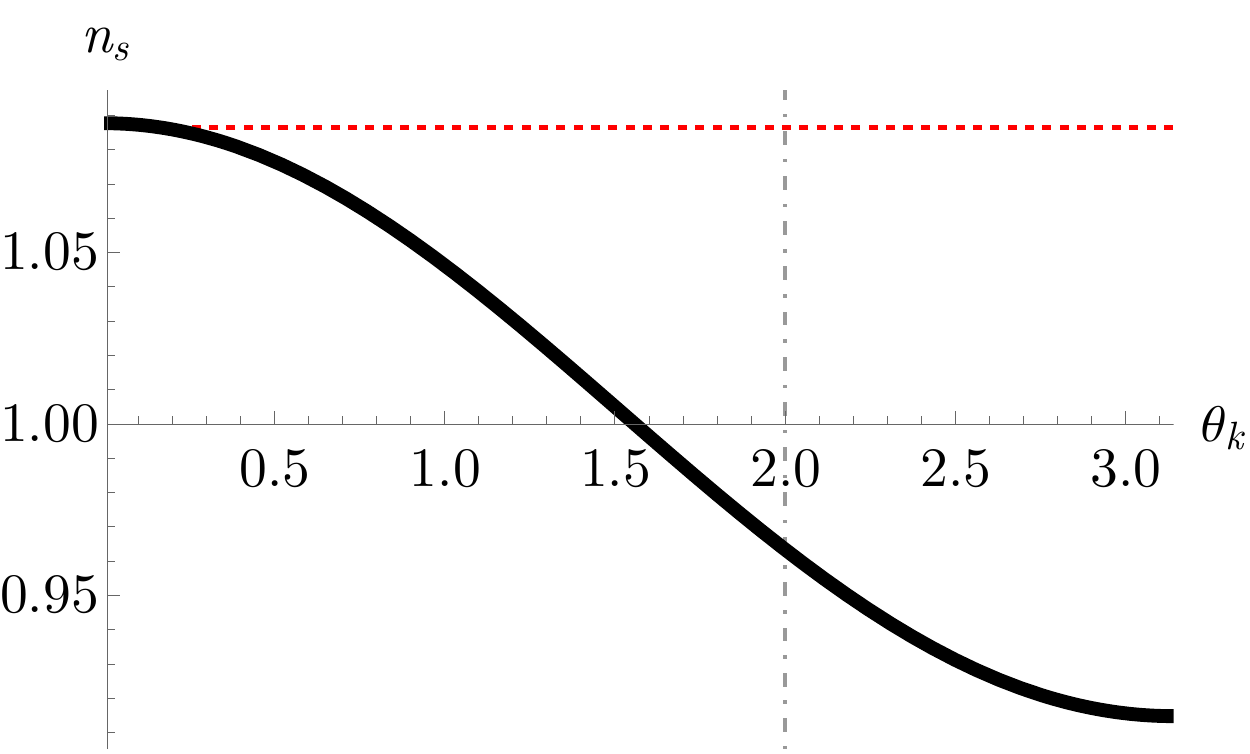}
  \end{center}
 \end{minipage} 
%
%
 \begin{minipage}{.32\linewidth}
  \begin{center}
 \includegraphics[width=\linewidth]{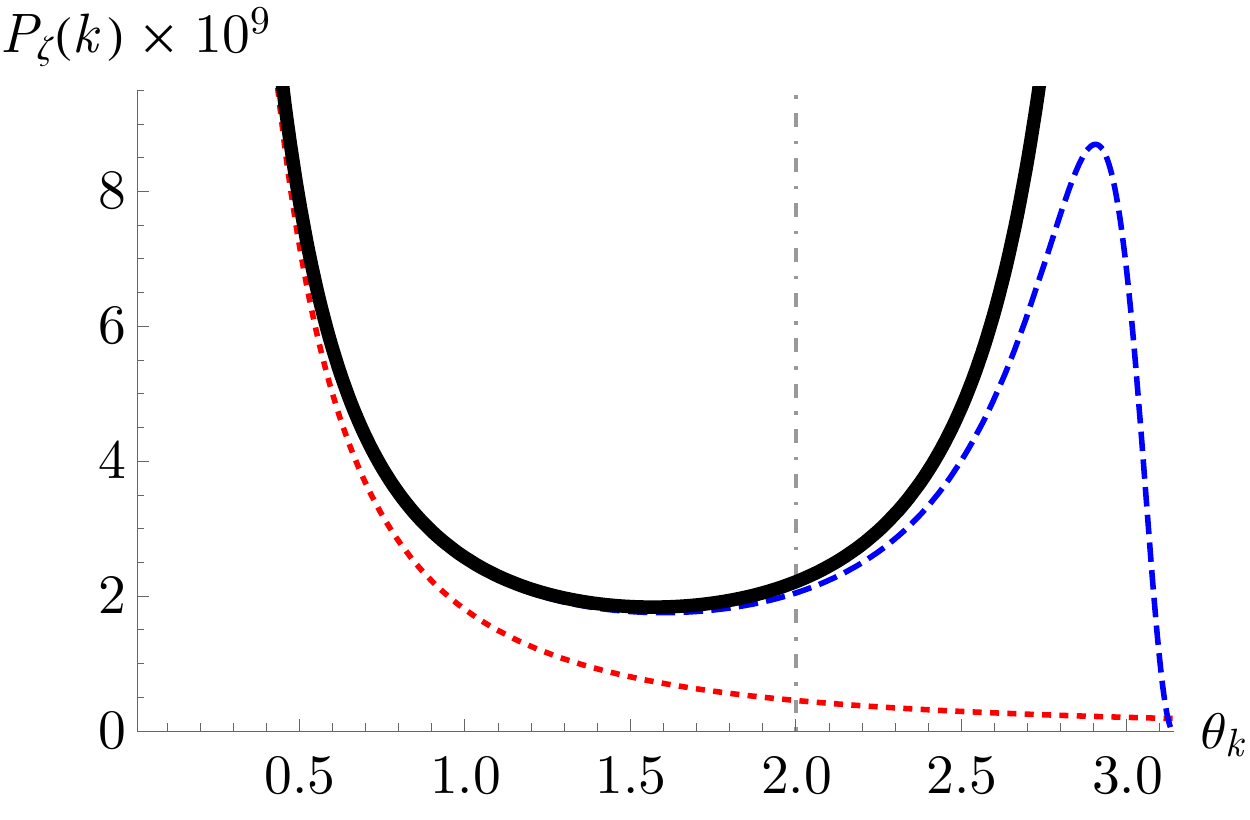}
  \end{center}
 \end{minipage} 
 \begin{minipage}{0.01\linewidth} 
  \begin{center}
  \end{center}
 \end{minipage} 
 \begin{minipage}{.32\linewidth}
  \begin{center}
 \includegraphics[width=\linewidth]{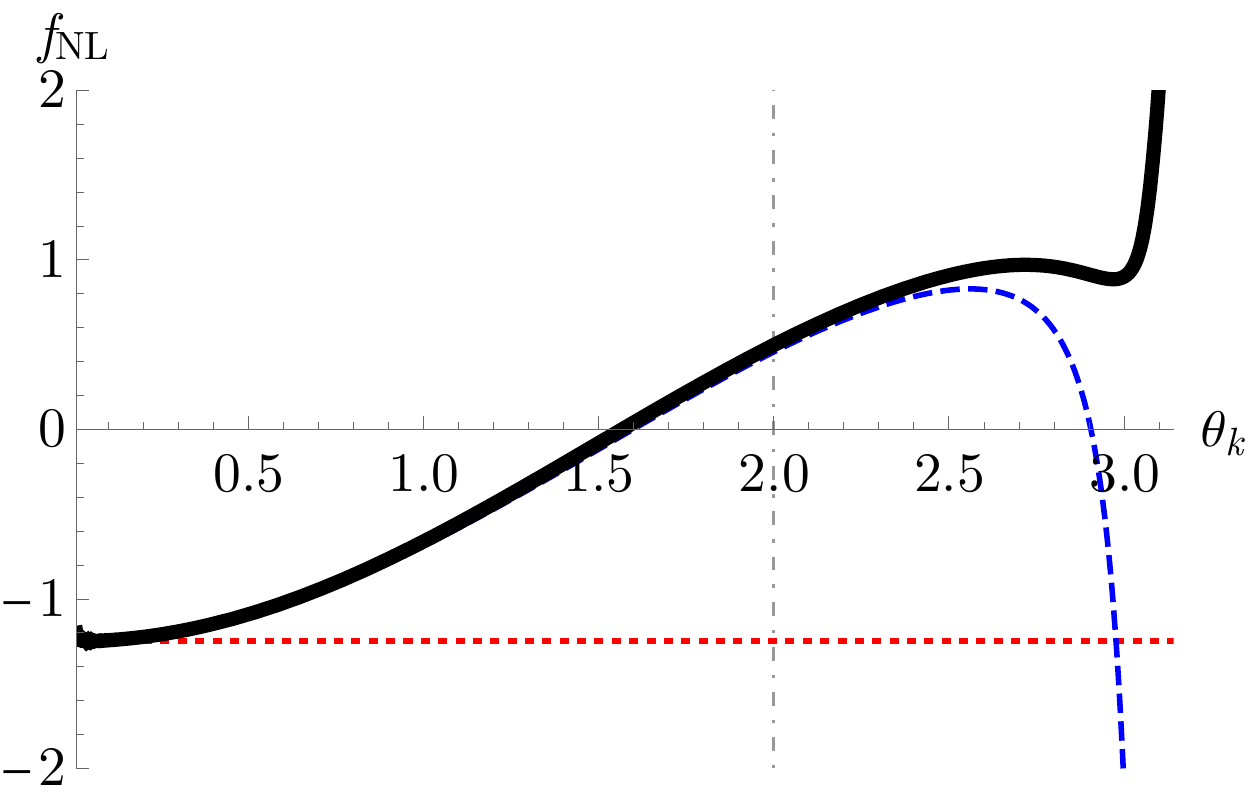}
  \end{center}
 \end{minipage} 
 \caption{Curvature perturbation as a
 function of the axion angle when the pivot scale exits the horizon
 during inflation. The axion decay constant and strong coupling scale
 are taken as $f = 2500 H_{\mathrm{inf}}$ and $\Lambda = 30
 H_{\mathrm{inf}}$, so that the power spectrum amplitude and the
 spectral index match with observations at $\theta_k \approx 2.0$. 
The black solid lines show the numerical results, while the blue dashed
 lines are analytic approximations. Results for vanilla curvatons
 are shown as the red dotted lines.} 
 \label{fig:curv}
\end{figure*}

In Fig.~\ref{fig:curv} we plot $n_s$, $P_\zeta$, and $f_{\mathrm{NL}}$
as functions of~$\theta_k$. Here we have fixed the axion parameters as 
$f = 2500 H_{\mathrm{inf}}$ and $\Lambda = 30  H_{\mathrm{inf}}$, 
with a sufficiently small photon coupling~$\alpha$ such that 
$ R \gg 1$ in the  entire displayed region of 
$10^{-2} \leq \theta_k < \pi  $. 
The mode $k$ exits the horizon about $50$ $e$-folds
before the end of inflation, 
and the time-variation of the inflationary Hubble rate is taken as
$\abs{\dot{H}/H^2} \ll 10^{-2}$ so that its effect on the spectral index
is negligible.
We note that the curvature perturbation is insensitive to the precise
values of $H_{\mathrm{inf}}$, $\alpha$, $p$, and $g_*$,
as long as the aforementioned assumptions are satisfied.
The black solid lines show the exact results obtained by numerically
solving the equation of motion of the axion and computing $\delta
\mathcal{N}$.
We have verified that the numerical results do not change if the axion
and inflaton decay suddenly or smoothly, 
as long as $ R \gg 1$ and also the inflaton decay width is 
not much smaller than $H_{\mathrm{inf}}$. 
The blue dashed lines show the $R \to \infty$ limit of the 
analytic expressions~(\ref{eq:Pzeta-fNL}),
and the red dotted lines show the results for a quadratic curvaton 
(i.e. $\theta_k \to 0$ and $R \to \infty$).
The analytic expression~(\ref{eq:n_s-1}) for the spectral index fully
overlaps with the numerical result, hence is not shown in the plot.
The behavior of the axion-induced perturbation reduces to that of a
quadratic curvaton at $\theta_k \ll 1$.
The analytic expressions in (\ref{eq:Pzeta-fNL})
are seen to match well with the numerical results
up to $\theta_k \lesssim 2.5$, which corresponds to 
$\theta_{\mathrm{end}} \lesssim 0.5$. 
For larger $\theta_k$, the inhomogeneous onset of the axion
oscillation becomes relevant and thus the approximation of constant~$c$
used for deriving (\ref{eq:Pzeta-fNL}) breaks down;
in particular, $f_{\mathrm{NL}}$ becomes of order~$10$
as $\theta_k $ approaches $\pi$ \cite{Kawasaki:2011pd}.
The Planck best-fit values 
$P_\zeta (k_{\mathrm{p}}) \approx 2.1 \times 10^{-9}$
and $n_s \approx 0.96$ at the pivot scale 
$k_{\mathrm{p}} = 0.05 \, \mathrm{Mpc}^{-1}$~\cite{Aghanim:2018eyx}
are realized at $\theta_k \approx 2.0$. 
Here the local-type non-Gaussianity is also within the Planck 68\% limits
$f_{\mathrm{NL}} = -0.9 \pm 5.1$~\cite{Akrami:2019izv}.

\section{Observational constraints}

We now investigate the parameter space.
The periodic axion potential comes equipped with 
negatively curved regions where the red-tilted perturbation can be sourced
without relying on large-field inflation.
The observed value of $n_s $ indicates
that the axion angle lies
within $\pi / 2 < \abs{\theta_k} < \pi $
when the cosmological scales exit the horizon,
and also that the zero-temperature axion mass is of
order $10^{-1} H_{\mathrm{inf}}$~\cite{max-e-folds}.
Current limits on non-Gaussianity favor $ R \gg 1$
(i.e. axion domination before decay), 
and also $\abs{\theta_k}$ not too close to $\pi$ where
the inhomogeneous onset of the oscillation enhances $f_{\mathrm{NL}}$.
Hence, supposing the axion to roll down to the region
$ \abs{\theta_{\mathrm{end}}} < \pi/2$ by the end of inflation,
we take
$-\cos \theta_k \sim \abs{\sin \theta_k} \sim \cos \theta_{\mathrm{end}}
\sim 1$
to make a rough estimate of the axion parameters.
Then (\ref{eq:Pzeta-fNL}) and (\ref{eq:n_s-1}) give
$P_\zeta \sim (H_{\mathrm{inf}}/3 \pi f)^2$,
$n_s - 1 \sim -2 m_0^2/3 H_{\mathrm{inf}}^2$,
and
$f_{\mathrm{NL}} \sim 5/4$.
Thus we find that the observed values of $P_\zeta$ and $n_s$ fix the decay
constant and strong coupling scale in terms of the inflation scale as
\begin{equation}
 f \sim 2000 H_{\mathrm{inf}},
\quad
 \Lambda \sim 20 H_{\mathrm{inf}}.
\label{eq:f_Lambda}
\end{equation}

One can then check, taking $c \sim 1 $, that most of the assumptions
in the previous sections are automatically satisfied as long as
the inflation scale does not exceed the observational upper bound
$H_{\mathrm{inf}} \lesssim 10^{14}\, \mathrm{GeV}$
\cite{Aghanim:2018eyx}.
On the other hand, the condition of 
$T_{\mathrm{max}} < f$ (cf. (\ref{eq:i})),
without which the axion would temporarily vanish after inflation and
thus would 
not be able to source the perturbation,
imposes a lower bound on the inflation scale as 
\begin{equation}
 H_{\mathrm{inf}} \gtrsim 10^{11}\, \mathrm{GeV}.
\label{eq:Vtilde}
\end{equation}
Hence our scenario is compatible not only with high scale inflation
producing observably large primordial gravitational waves, but also with
inflation happening at lower scales.
We remark that the lower bound (\ref{eq:Vtilde}) is further relaxed if 
the inflaton does not decay instantaneously and yields
$T_{\mathrm{max}}$ lower than (\ref{eq:T_max}), 
or if the symmetry breaking scale is larger than the axion
periodicity (possibly with a large domain wall number).

The decay width of the axion is constrained by the 
assumption of $ R \gg 1$, 
and also from the requirement that the cosmic temperature after the
decay be higher than $\sim 4\, \mathrm{MeV}$ so as not to spoil Big Bang
Nucleosynthesis~\cite{Kawasaki:2000en}. 
The former condition restricts the 
axion-photon coupling~$\alpha$ from above, and the latter from below.
These bounds become more stringent for lower inflation scales;
for $H_{\mathrm{inf}} = 10^{11} \, (10^{14}) \, \mathrm{GeV}$ 
the coupling is bounded as  
\begin{equation}
 10^{-11} \, (10^{-13}) \lesssim \alpha \lesssim 10^{-2} \, (10^3).
\end{equation}
We also note that if $\alpha$ greatly exceeds unity then
the effective theory would break down at $T = T_{\mathrm{max}}$.

\section{Discussion}

We showed that an axionlike particle coupled to a new confining
sector can generate the primordial density perturbation of our universe.
The axion decay constant and strong coupling scale are
uniquely determined by observations in terms of the inflation scale
as (\ref{eq:f_Lambda}), which, in the minimal model, 
entail a temporal deconfinement of the gauge group after inflation.

Let us comment on the observational consequences.
The axionlike scenario predicts a {\it positive}
local-type non-Gaussianity of order unity, $f_{\mathrm{NL}} \sim 1$,
on large scales.
This is in contrast to single-field inflation which yields a much
smaller local $f_{\mathrm{NL}}$, 
and also to a vanilla curvaton which produces
$f_{\mathrm{NL}} = -5/4 $ in the dominating limit ($R \to \infty$).
These values of non-Gaussianity are within reach of upcoming large-scale
structure surveys~\cite{Alvarez:2014vva}.
We also remark that the axion 
lies within $\pi / 2 < \abs{\theta} < \pi$ and sources the red spectral
tilt when the cosmological scales exit the horizon, 
and then rolls down typically to the region $\abs{\theta} < \pi / 2$ by
the end of inflation. 
This implies that the density perturbation spectrum becomes blue-tilted
at small scales and thus is enhanced.
Likewise, $f_{\mathrm{NL}}$ runs towards negative values at small scales.
These features could be tested observationally using, for instance, 
ultracompact minihalos~\cite{Bringmann:2011ut}.

\section*{Acknowledgments}

I am grateful for discussions with Yuichiro Tada, Hiroyuki Tashiro,
 Lorenzo Ubaldi, Matteo Viel, and Shuichiro Yokoyama.




\begin{thebibliography}{99}

\bibitem{Peccei:1977hh}
R.~Peccei and H.~R.~Quinn,
Phys.\ Rev.\ Lett.\  \textbf{38}, 1440-1443 (1977);
%
S.~Weinberg,
Phys.\ Rev.\ Lett.\  \textbf{40}, 223-226 (1978);
%
F.~Wilczek,
Phys.\ Rev.\ Lett.\  \textbf{40}, 279-282 (1978).

\bibitem{Svrcek:2006yi}
P.~Svrcek and E.~Witten,
JHEP \textbf{06}, 051 (2006)
[arXiv:hep-th/0605206 [hep-th]];
%
M.~R.~Douglas and S.~Kachru,
Rev.\ Mod.\ Phys.\  \textbf{79}, 733-796 (2007)
[arXiv:hep-th/0610102 [hep-th]];
%
A.~Arvanitaki, S.~Dimopoulos, S.~Dubovsky, N.~Kaloper and J.~March-Russell,
Phys.\ Rev.\ D \textbf{81}, 123530 (2010)
[arXiv:0905.4720 [hep-th]].

\bibitem{Mollerach:1989hu}
S.~Mollerach,
Phys.\ Rev.\ D \textbf{42}, 313-325 (1990);
%
A.~D.~Linde and V.~F.~Mukhanov,
Phys.\ Rev.\ D \textbf{56}, 535-539 (1997)
[arXiv:astro-ph/9610219 [astro-ph]];
%
K.~Enqvist and M.~S.~Sloth,
Nucl.\ Phys.\ B \textbf{626}, 395-409 (2002)
[arXiv:hep-ph/0109214 [hep-ph]];
%
D.~H.~Lyth and D.~Wands,
Phys.\ Lett.\ B \textbf{524}, 5-14 (2002)
[arXiv:hep-ph/0110002 [hep-ph]];
%
T.~Moroi and T.~Takahashi,
Phys.\ Lett.\ B \textbf{522}, 215-221 (2001)
[arXiv:hep-ph/0110096 [hep-ph]].

\bibitem{Dimopoulos:2003az}
K.~Dimopoulos, D.~Lyth, A.~Notari and A.~Riotto,
JHEP \textbf{07}, 053 (2003)
[arXiv:hep-ph/0304050 [hep-ph]].

\bibitem{Kawasaki:2011pd}
M.~Kawasaki, T.~Kobayashi and F.~Takahashi,
Phys.\ Rev.\ D \textbf{84}, 123506 (2011)
[arXiv:1107.6011 [astro-ph.CO]];
%
M.~Kawasaki, T.~Kobayashi and F.~Takahashi,
JCAP \textbf{03}, 016 (2013)
[arXiv:1210.6595 [astro-ph.CO]].

\bibitem{quadratic_c}
$ c =   (2p+4) [ \Gamma (\frac{2 p+5}{2 p+4} ) / \sqrt{\pi}
]^{\frac{2 p+4}{p+3}}$
for a quadratic potential with a temperature-dependent mass
of the form (\ref{eq:mofT});  
see 
T.~Kobayashi and L.~Ubaldi,
[arXiv:2006.09389 [hep-ph]].

\bibitem{Starobinsky:1986fxa}
A.~A.~Starobinsky,
JETP Lett.\  \textbf{42}, 152-155 (1985);
%
M.~Sasaki and E.~D.~Stewart,
Prog.\ Theor.\ Phys.\  \textbf{95}, 71-78 (1996)
[arXiv:astro-ph/9507001 [astro-ph]];
%
D.~Wands, K.~A.~Malik, D.~H.~Lyth and A.~R.~Liddle,
Phys.\ Rev.\ D \textbf{62}, 043527 (2000)
[arXiv:astro-ph/0003278 [astro-ph]];
%
D.~H.~Lyth, K.~A.~Malik and M.~Sasaki,
JCAP \textbf{05}, 004 (2005)
[arXiv:astro-ph/0411220 [astro-ph]].

\bibitem{Aghanim:2018eyx}
N.~Aghanim \textit{et al.} [Planck],
[arXiv:1807.06209 [astro-ph.CO]].

\bibitem{Akrami:2019izv}
Y.~Akrami \textit{et al.} [Planck],
[arXiv:1905.05697 [astro-ph.CO]].

\bibitem{max-e-folds}
The requirement of $\abs{\theta_k} > \pi / 2$ constrains the duration of
inflation prior to horizon exit of the cosmological scales;
e.g., with the parameters in Fig.~\ref{fig:curv}, the axion
rolls from $\theta_i = 0.9 \, (0.99) \times \pi$ to $\theta_k =
2.0$ in about $30 \, (90)$ $e$-folds. 
However if the axion evolution is dominated by quantum
fluctuations during the early stage of inflation (which happens
in this example if $\abs{\theta_i} \gtrsim 0.9995 \times \pi$),
then inflation could last much longer. 

\bibitem{Kawasaki:2000en}
M.~Kawasaki, K.~Kohri and N.~Sugiyama,
Phys.\ Rev.\ D \textbf{62}, 023506 (2000)
[arXiv:astro-ph/0002127 [astro-ph]];
%
S.~Hannestad,
Phys.\ Rev.\ D \textbf{70}, 043506 (2004)
[arXiv:astro-ph/0403291 [astro-ph]].

\bibitem{Alvarez:2014vva}
M.~Alvarez \textit{et al.},
[arXiv:1412.4671 [astro-ph.CO]].

\bibitem{Bringmann:2011ut}
T.~Bringmann, P.~Scott and Y.~Akrami,
Phys. Rev. D \textbf{85}, 125027 (2012)
[arXiv:1110.2484 [astro-ph.CO]];
%
M.~S.~Delos, A.~L.~Erickcek, A.~P.~Bailey and M.~A.~Alvarez,
Phys. Rev. D \textbf{98}, no.6, 063527 (2018)
[arXiv:1806.07389 [astro-ph.CO]];
%
K.~Furugori, K.~Abe, T.~Tanaka, D.~Hashimoto, H.~Tashiro and K.~Hasegawa,
[arXiv:2002.04817 [astro-ph.CO]].
	
\end{thebibliography}
\end{document}